\documentclass[useAMS,usenatbib,usegraphicx]{mn2e}

\title[Structure of Brightest Cluster Galaxies]{Environmental Dependence of the Structure of Brightest Cluster Galaxies}
\author[Brough et al.]
       {S.~Brough,$^{1,2}$\thanks{Email: sbrough@astro.swin.edu.au} C. A.~Collins,$^1$ D. J.~Burke,$^3$ P. D.~Lynam$^4$, R. G.~Mann$^5$\\$^1$Astrophysics Research Institute, Liverpool John Moores University, Egerton Wharf, Birkenhead, CH41 1LD, UK\\$^2$Centre for Astrophysics and Supercomputing, Swinburne University of Technology, Hawthorn, VIC 3122, Australia\\$^3$Harvard-Smithsonian Center for Astrophysics, 60 Garden Street, Cambridge, MA 02138, USA\\$^4$European Southern Observatory, Casilla 19001, Santiago 19, Chile\\$^5$Institute for Astronomy, University of Edinburgh, Royal Observatory, Blackford Hill, Edinburgh, EH9 3NJ, UK}

\begin{document}

\date{Accepted... Received...; in original form 2005}

\pagerange{\pageref{firstpage}--\pageref{lastpage}} \pubyear{2005}

\maketitle

\label{firstpage}

\begin{abstract}
We measure the Petrosian structural properties of 33 brightest cluster
galaxies (BCGs) at redshifts $z\leq0.1$ in X-ray selected clusters
with a wide range of X-ray luminosities.  We find that some BCGs show
distinct signatures in their Petrosian profiles, likely to be due to
cD haloes.  We also find that BCGs in high X-ray luminosity clusters
have shallower surface brightness profiles than those in low X-ray
luminosity clusters.  This suggests that the BCGs in high X-ray
luminosity clusters have undergone up to twice as many equal-mass
mergers in their past as those in low X-ray luminosity clusters.  This
is qualitatively consistent with the predictions of hierarchical
structure formation.
\end{abstract}


\section{Introduction}
\label{intro}

The first-ranked or brightest cluster galaxies (BCGs) are uniquely
positioned at the centre of the gravitational potential of clusters of
galaxies.  With brightnesses roughly ten times that of $L^\star$
galaxies, BCGs are not simply the bright extension of a Schechter
luminosity function (e.g. \citealt{s76,tremaine77,bernstein01}).  They
are predominantly elliptical in their morphology and owing to their
extreme luminosities and position, they dominate their host cluster
visually, forming a unique population in their own right.  However,
their evolution and how their environment affects that evolution, are
still poorly understood.

In previous papers of this series on the K-band photometric properties
of BCGs; \cite{cm98}; \cite{bcm} and \cite{brough02}: we observe that
BCGs in clusters with an X-ray luminosity $L_{X} > 1.9\times10^{44}$
erg s$^{-1}$ (in the EMSS passband $0.3-3.5$ keV with
$\Omega_{\Lambda}=0.7$, $\Omega_m=0.3, h=0.7$) are brighter and more
uniform than those in their low X-ray luminosity counterparts at
redshifts $z \geq 0.1$.  It appears from \cite{brough02} that the
evolutionary history of BCGs depends on the richness of their cluster
environment, with BCGs in the most luminous clusters showing little
sign of merging since $z\sim1$ and those in the least luminous
clusters showing significant signs of mass growth over the same
timescale.  However, the mass accretion rates inferred in
\cite{brough02} rely on the assumption that increases in luminosity
are attributable to increases in mass.

It is important to confirm this result to determine whether the mass
evolution of BCGs is dependent on their environment.  If BCGs are in
dynamical equilibrium then changes in mass should produce changes in
size.  Therefore, a means to test the assumption of increases in
luminosity being attributable to mass accretion, and confirm the
results from \cite{brough02}, is to measure the structural parameters
(size and surface brightness) of BCGs with respect to the properties
of their host cluster environment.

Many authors have studied the effects of mergers on galaxy surface
brightness profiles since \cite{toomre72} suggested that dynamical
friction, and the consequent merging, of pairs of disc galaxies could
form elliptical galaxies.  Numerical simulations indicate that the de
Vaucouleurs $r^{1\over{4}}$ surface brightness profile \citep{dev} is
robust to the effects of mergers \citep{barnes88,navarro90}.
The profile of the remnant is only disturbed with respect to the de
Vaucouleurs prediction at radii greater than the effective radius and
these disturbances move outwards with time, disappearing within the
dynamical timescale of the galaxy.  \cite{schweizer82} observed that
the $V$-band light distribution of NGC7252, a galaxy which appears to
have undergone a recent merger, had already established an
$r^{1\over{4}}$ profile.  \cite{wright90,james99} and
\cite{brown03} examined $K$-band surface brightness profiles of merger 
galaxies and merger remnants, all confirming that the structure of the
mergers more closely resembled that of elliptical galaxies than spiral
bulges.

More recently, mergers between elliptical galaxies have been simulated
in order to understand the observed tilt of the Fundamental Plane (FP;
between radius, surface brightness and velocity dispersion) with
respect to the Virial theorem.  These simulations find that the
effective radius of merger remnants increases with total mass --
(e.g. \citealt{navarro90,capaccioli92,capelato95,dantas03,evstigneeva04}),
with the surface brightness, consequently, decreasing.
Equal-mass dissipationless merging of simulated elliptical galaxies
produces structurally non-homologous remnants and is, therefore, able
to reproduce the observed FP \citep{capelato95,nipoti03}.  
These studies suggest that galaxies become larger and more diffuse as
a function of merger events.



Observations of the Kormendy relation (between radius and surface
brightness; \citealt{kormendy}) of BCGs show that they lie off the
relationship followed by normal cluster elliptical galaxies, having
larger radii and lower mean surface brightnesses than is predicted by
simply scaling the relationships followed by normal elliptical
galaxies \citep{thuan81,hoessel87,schombert2,bcgfp}.  However, they do
lie on the same FP as normal cluster ellipticals
\citep{bcgfp}.  These observations are consistent with the suggestion
that BCGs have undergone an increased merger history with respect to
normal cluster elliptical galaxies.  There are also suggestions that
the de Vaucouleurs effective radii of BCGs correlate with their
environment.  \cite{garilli97} fitted de Vaucouleurs $r^{1\over{4}}$
laws to the surface brightness profiles of 17 BCGs with redshifts
$z<0.2$ in the $g-$, $r-$ and $i-$ bands.  They find a correlation
between effective radius and local galaxy density (measured as the
number of galaxies brighter than $M_V=-18$ in circles of projected
radii equal to 200 kpc) in the sense that larger BCGs are found in
environments which are locally denser at the present epoch.  Few other
studies have taken environment into account so it is unclear what role
environment plays in the structural differences observed between
normal cluster elliptical galaxies and BCGs.


In this paper the structural properties of BCGs are measured to
further examine how the properties of BCGs depend on their host
cluster environment, with the aim of interpreting their evolutionary
history.
This analysis uses a larger sample of objects covering a significantly
wider range in environmental density than previous studies.  An
objective measure of environmental density is vitally important in any
study of BCG properties.  As the cluster X-ray luminosity is
proportional to the volume integral of the square of the density of
the intra-cluster medium it provides a quantitative measure of the
environmental density of BCGs with which to analyse possible
environmental dependences.  We also make use of near-infrared imaging
with its associated benefits of small evolutionary corrections and low
Galactic extinction

The sample is outlined in Section~\ref{sample_4} along with a
description of the reduction of the data.  The analysis techniques
are presented in Section~\ref{analysis_4}.
The results obtained are introduced in Section~\ref{pet_kormendy} and
discussed in Section~\ref{discuss_4} and the conclusions drawn from
this paper are presented in Section~\ref{summary_4}.

\section{Data}
\label{sample_4}

\subsection{Sample}

Following on from the photometric analysis in \cite{brough02} the aim
was to construct a sample of BCGs in X-ray selected clusters with a
wide range of X-ray luminosities.  Structural analysis requires high
resolution images and to achieve this our analysis is primarily
concerned with the structural properties of BCGs at redshifts
$z\leq0.1$.  

The \cite{bcm} sample consists of 78 BCGs with redshifts $0.0<z<0.8$
in clusters selected from Einstein Medium-Sensitivity Survey (EMSS;
\citealt{gl94}) and the Serendipitous High-redshift Archival ROSAT
Cluster (SHARC; \citealt{b97,burke03,r00}) X-ray cluster catalogues.
These galaxies were observed in the $K$-band with the Infra-Red CAMera
3 (IRCAM3) on the United Kingdom Infra-Red Telescope (henceforth
UKIRT; 1 pixel = 0.281$^{\prime\prime}$ giving a field of view of
72$^{\prime\prime}$) between 1994 and 1998.  Eleven of these BCGs are
at redshifts $z<0.1$ (all from the EMSS catalogue) and, therefore,
suitable for use in this analysis.

In contrast to the analysis in \cite{brough02} it was not possible to
use the 2MASS extended source catalogue \citep{2mass} for this
analysis.  2MASS only provides one surface brightness measurement and
the two images it provides are not adequate for a structural analysis:
The postage stamp images are scaled to the size of the galaxy and are
too small to analyse to a reasonable radius.  The images from the
actual scans, have a scale of 1$^{\prime\prime}$/pixel which is too
large to provide a meaningful analysis.

Therefore, to extend this analysis a further 26 BCGs with redshifts
$0.04<z<0.1$ selected from the Lynam catalogue \citep{pdl} were
observed over 3 nights (14-16 January 2001) in the K$_n$-band with the
QUick Infra-Red Camera (QUIRC) on the University of Hawaii 2.2m
telescope (henceforth UH; 1 pixel = 0.189$^{\prime\prime}$ giving a
field-of-view of 193$^{\prime\prime}$).  These BCGs are in clusters
selected from the NOrthern {\it ROSAT} All-Sky X-ray cluster survey
(NORAS; \citealt{boh00}).  The UKIRT and UH data were reduced in the
same manner and a brief overview of this process is given below.

As there are two galaxies in common between the UH and UKIRT samples
(those in the clusters MS0301-7 and MS0904-5) this gives a total
sample of 35 BCGs, all with redshifts $z\leq0.1$.

\subsection{Data Reduction}
The data reduction was performed using the {\sc IRAF/DIMSUM} (Deep
Infrared Mosaicing Software UM) package.

The individual frames were masked for bad pixels, dark subtracted and divided 
by the exposure time.  A flat field image was created by median combination of 
the object images, or separate sky exposures if these were available, and 
applied to the object frames.  Masking of cosmic ray events was performed on 
the flattened images before they were mosaiced together, which completed the 
processing of those objects with sky exposures.  Otherwise  the mosaic, which 
is substantially deeper than the individual exposures, was used to create an 
object mask, which was then applied to the individual images before they were 
median-combined to form a flat.  The flattened images were then processed as 
above to create the final image.

Observations of stars from the UKIRT faint standards list
\citep{casali92} were used to calibrate the UKIRT photometry onto the UKIRT
system assuming an extinction of $0.088$ mag airmass$^{-1}$, the
median value for K-band observations at Mauna Kea.

The UH images were calibrated using observations of standard stars
selected from the UKIRT extended faint standards list
\citep{hawarden01}.
At least 7 different standards were observed at least 3 times each
night.  With little variation over the 3 nights, the observations were
combined to give an atmospheric extinction coefficient of $0.118$ mag
airmass$^{-1}$ and a zero-point of $23.305$ with a standard deviation
$\sigma=0.029$ mag.

\section{Analysis}
\label{analysis_4}

One of the major problems in measuring the structural properties of
any galaxy type is systematic errors in the measurement of the sky
levels (e.g. \citealt{graham96,gonzalez01}) as even the smallest
excess flux can swamp the galaxy flux in the outer annuli, and
oversubtraction results in a truncation of the profile.  This problem
is increased when analysing BCGs, due to their larger size and the
fact that, locally, BCGs are observed to have significant amounts of
flux in their outer wings \citep{graham96}.

In this analysis the sky levels were calculated using SExtractor
\citep{sext}.  In this program the sky level is estimated by
constructing a grid and estimating the background within each square
with a sigma-clipped mean.  The grid is then median filtered to remove
the effects of large bright objects.  A grid size of 64 pixels and a
median filter size of $3\times3$ grid squares was used here in order
to avoid the background estimation being affected by the presence of
other galaxies.  This method was compared with measuring the
background by obtaining the median of the counts measured in small
annuli at different points around the image, the difference was of the
order of 10 per cent in $\sigma_{background}$ calculated by SExtractor
and no systematic offsets were observed.

Owing to their position at the centres of clusters of galaxies, BCGs
are often surrounded by many other galaxies.  It is important not to
incorporate the light of these galaxies when measuring the surface
brightness profile of the BCG.  In order to avoid this problem, the
flux from the neighbouring galaxies needs to be carefully excluded
from the images. When fitting surface brightness profiles in 2D it is
possible to deconvolve each galaxy and subtract the flux due to the
curves of growth of the companion galaxies, replacing the actual BCG
flux in each excluded pixel with that calculated using its symmetric
counterpart about the centre of the BCG
(e.g. \citealt{lauer86,garilli97,nelsonc}).  However, this assumes
that the BCG is symmetric and that there is not a further object
affecting the counterpart position.

To reduce systematic uncertainties in this analysis these assumptions
were not made.  Pixels that were identified as belonging to a
contaminating source were excluded from the profile calculation. The
exclusion radius around each source was initially set to twice the
semi-major axis of the contaminating source (as reported by
SExtractor). These apertures were then inspected by eye, and the
exclusion radii of those sources close to the BCG centre were reduced,
since the BCG flux close to the centre is so high that flux from other
galaxies is insignificant.


The galaxies were then fitted with elliptical isophotes using the
{\small IRAF} task {\small ELLIPSE} (\citealt{ellipse};
e.g. \citealt{garilli97,gonzalez01}) which calculates the {\it median}
intensity within each isophote, further reducing the effect flux from
companion galaxies can have on the profile of the BCG.  The centres of
the ellipses were fixed while the position angles and ellipticities
were allowed to vary freely.

{\small ELLIPSE} was allowed to run until either the angular extent of
an elliptical annulus reached the width of the image or greater than
75 per cent of the pixels in a single annulus were masked out.
Intensities differing from the median by more than 2 standard
deviations were excluded to ensure the quality of the data being
fitted.  The radii were then circularized ($r=\sqrt{ab}$) and
converted from arcseconds to $h^{-1}$ kpc assuming a WMAP cosmology
\citep{spergel03}: $\Omega_m=0.3$, $\Omega_{\Lambda}=0.7$ and $H_0=100
h$ km s$^{-1}$ Mpc$^{-1}$ with $h=0.7$.

\subsection{Petrosian Structural Parameters}

Petrosian structural parameters (\citealt{petrosian76}) have a number
of advantages over those calculated from fits of models (e.g. de
Vaucouleurs law, S\'{e}rsic law) to the surface brightness profile.
In particular, no assumption of the stellar light distribution is made
in their calculation, making them a direct measure of galaxy
properties, from the local surface brightness.  The Petrosian
parameters are also relatively insensitive to zero-point errors and
errors in the extinction correction, as is clear from the Petrosian
relationship:
\begin{equation}
\eta(r)=\mu(r) - \langle \mu(r) \rangle,
\label{petr_eq2}
\end{equation}
where $\mu(r)$ is the surface brightness (in magnitudes) at a radius $r$ and 
$\langle \mu(r) \rangle$ is the mean surface brightness within $r$.  

\begin{figure*}
\begin{center}
    \resizebox{20pc}{!}{
     \rotatebox{-90}{
	\includegraphics{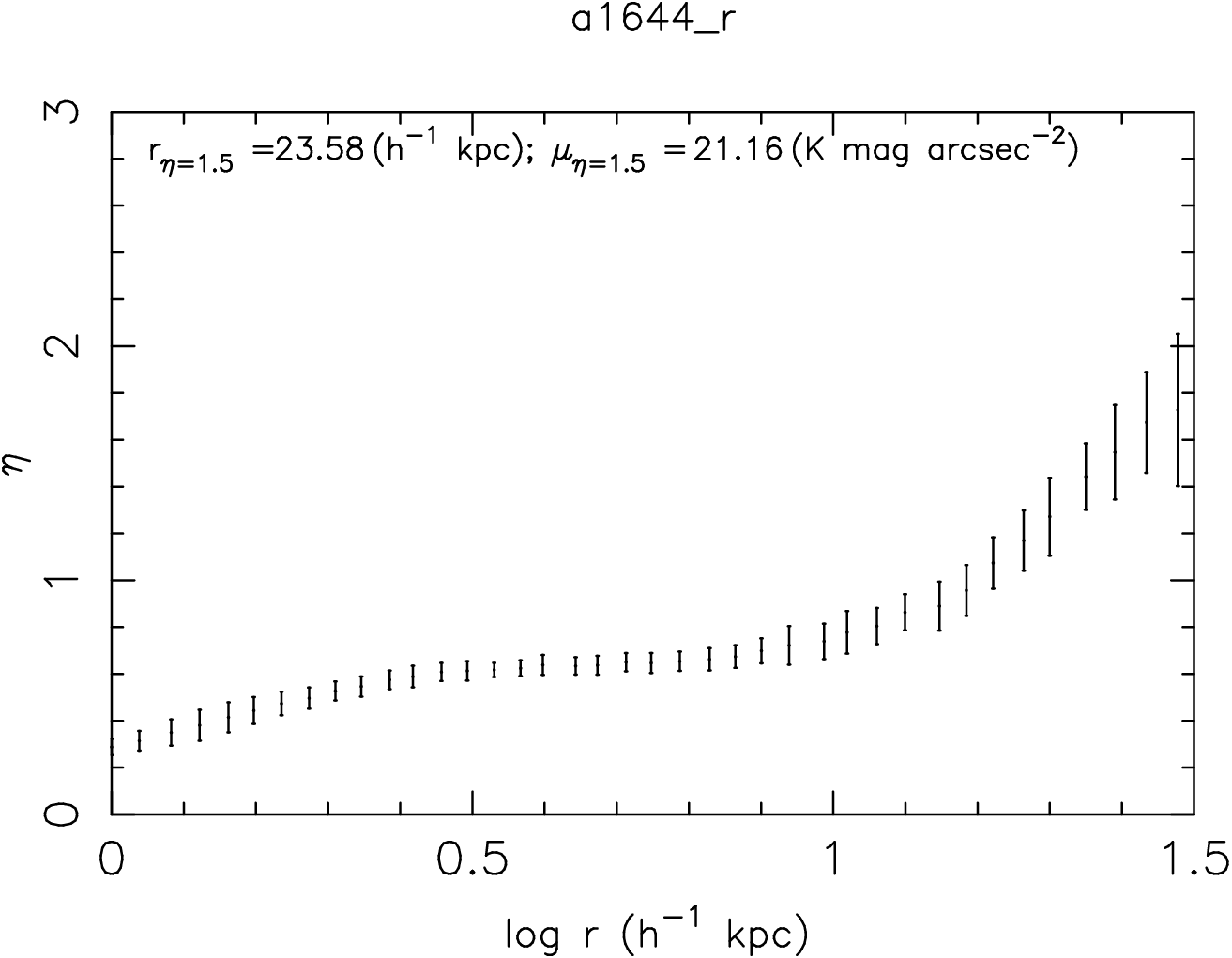}
	\includegraphics{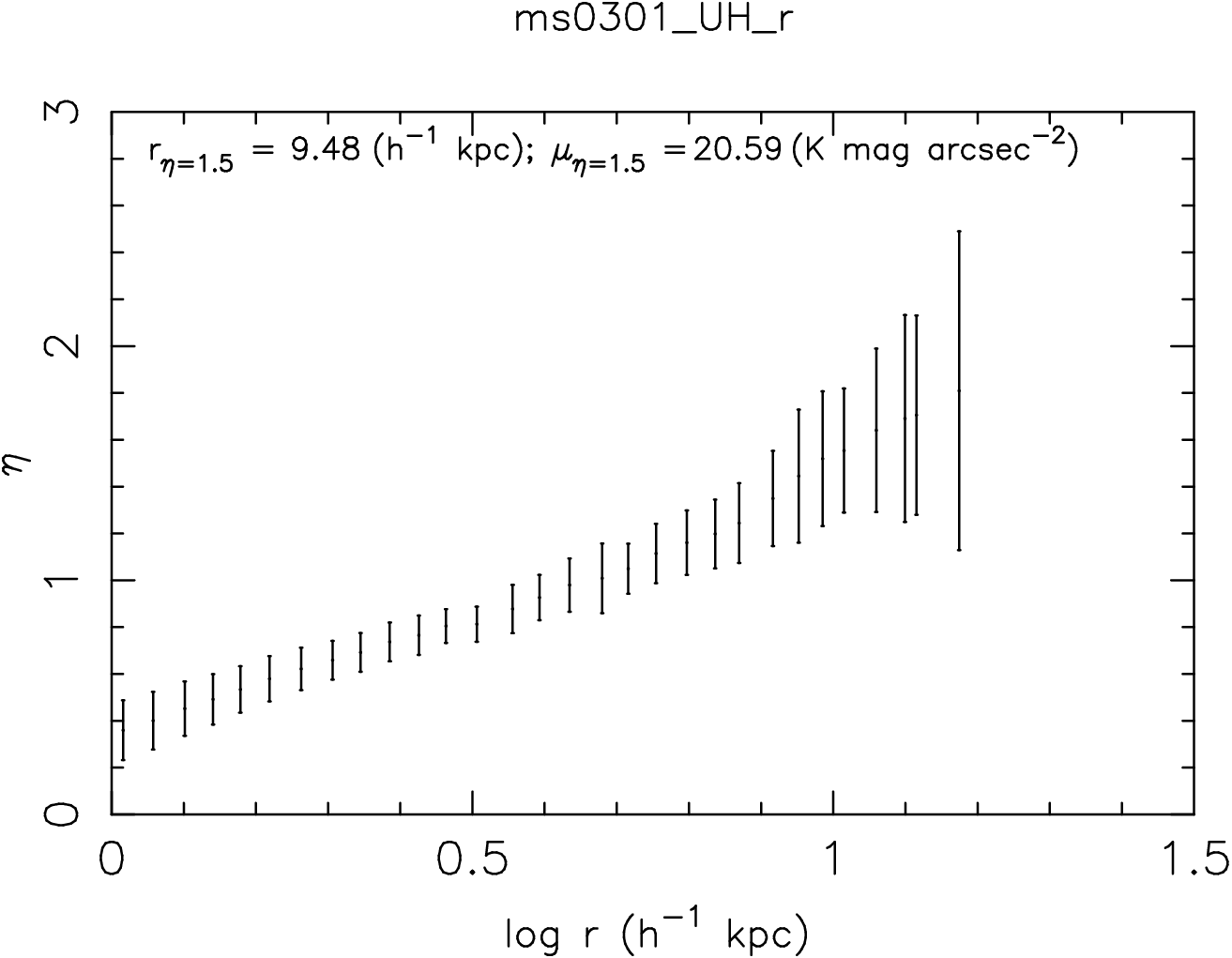}
}
}
    \resizebox{20pc}{!}{
      \rotatebox{-90}{
       \includegraphics{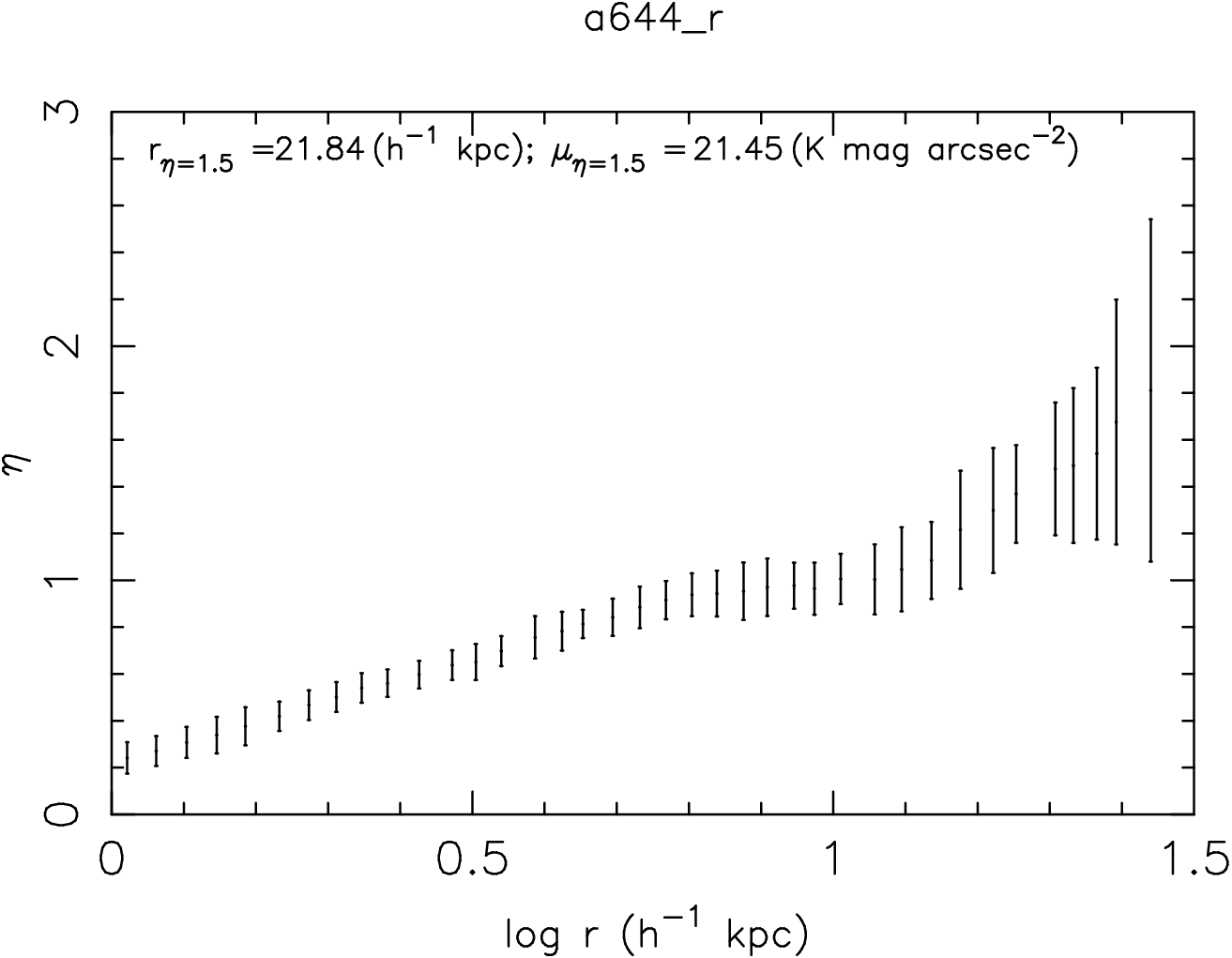}
	\includegraphics{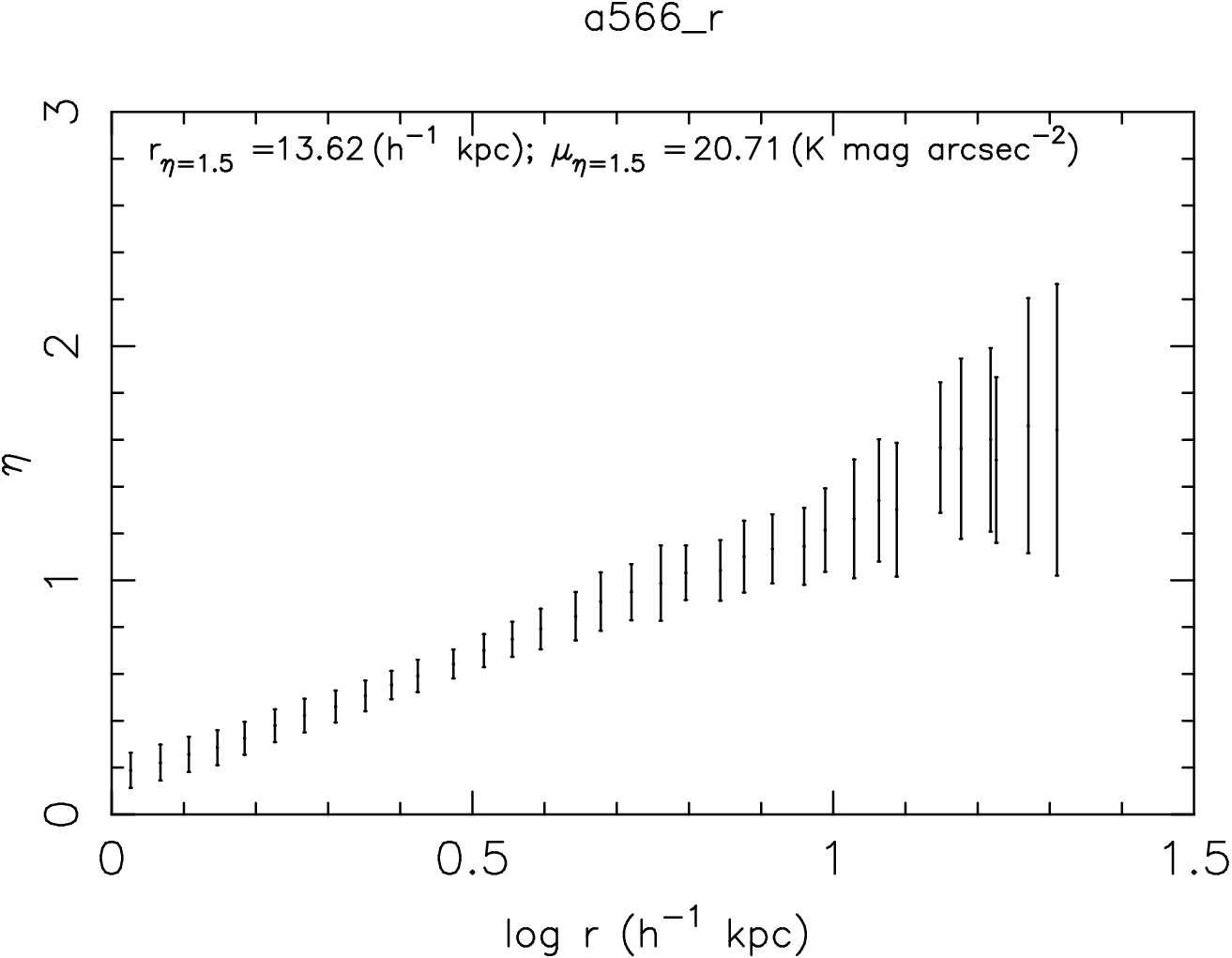}
}
}
\end{center}
\caption{Example $\eta(r)$ profiles of the BCGs in A1644, A644 (both 
high X-ray luminosity clusters), and those in MS0301-7 and A566 (both
low X-ray luminosity clusters).  The caption on each panel gives the
Petrosian radius, $r_ {\eta}$, and surface brightness, $\mu_{\eta}$,
measured from each profile at $\eta=1.5$.  The error bars indicate the
$1\sigma$ statistical errors on $\eta$ calculated by combining the
measured errors on $\mu(r)$ and $\langle \mu(r) \rangle$ in
quadrature.}
\label{eta_profile}
\end{figure*}

Figure~\ref{eta_profile} shows some example $\eta(r)$ profiles.  The
error bars indicate the $1\sigma$ statistical errors on $\eta$
calculated by combining the measured errors on $\mu(r)$ and $\langle
\mu(r) \rangle$ in quadrature.

\begin{figure}
\begin{center}

    \resizebox{20pc}{!}{
     \rotatebox{-90}{
	\includegraphics{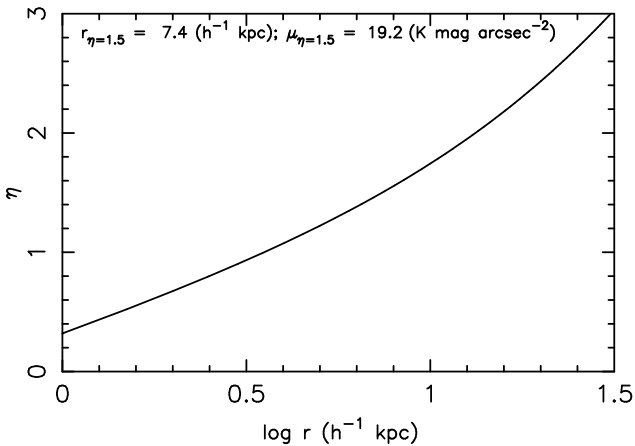}
    }
}
  \end{center}
\caption
{Example $\eta(r)$ profile for a model galaxy following a perfect de
Vaucouleurs law ($r_e=5~h^{-1}$ kpc; $\mu_e=17.8$ mag arcsec$^{-2}$).}
\label{kjaer_eta}
\end{figure}

The Petrosian radii and surface brightness can be measured from the
$\eta$ profile at a chosen value of $\eta$.  The choice of $\eta$
value to use depends on the sample: $r_{\eta}$ should be large enough
to be unaffected by seeing ($\eta\geq1.5$; \citealt{sandage90}) but
not so large that the surface brightnesses are noisy and unduly
sensitive to sky subtraction errors.  Figure~\ref{eta_profile}
illustrates that the errors on the $\eta$ parameter increase with
$\eta$, and \cite{kjaergaard93} find larger errors in their $r_{\eta}$
at $\eta=2$ than at $\eta=1.5$.  The images for this analysis have
seeing of $\sim1.5^{\prime\prime}$ and the images have typical
limiting depths of K$\sim22$ mag arcsec$^{-2}$, therefore, Petrosian
parameters calculated at $\eta=1.5$ satisfy both of these
requirements.
This results in a sample of BCGs with Petrosian radii in the range
$7^{\prime\prime}<r_{\eta=1.5}<37^{\prime\prime}$ and surface
brightnesses 19.5 mag arcsec$^{-2}<\mu_{\eta=1.5}<22.0$ mag
arcsec$^{-2}$.

The average Petrosian surface brightness $\langle \mu_{\eta} \rangle$ was 
calculated 
by summing the counts within $r_{\eta}$ and averaging over the area
enclosed.  $\langle \mu_{\eta} \rangle$ was then corrected for
cosmological dimming ($[1+z]^4$) and Galactic absorption using the
maps of \cite{ext98}; the absorption correction is small, typically
$\sim0.02$ mag.  The surface brightnesses were also K-corrected using
model predictions from the {\sc GISSEL96} code \citep{bc93}, following
\cite{yoshii88}.  These K-corrections also include an evolutionary 
correction, assuming pure luminosity evolution from a redshift of
formation, $z_f=2$.  The results are unchanged if the formation
redshift is increased to $z_f=5$.

\begin{figure}
\begin{center}
    \resizebox{20pc}{!}{
     \rotatebox{-90}{
	\includegraphics{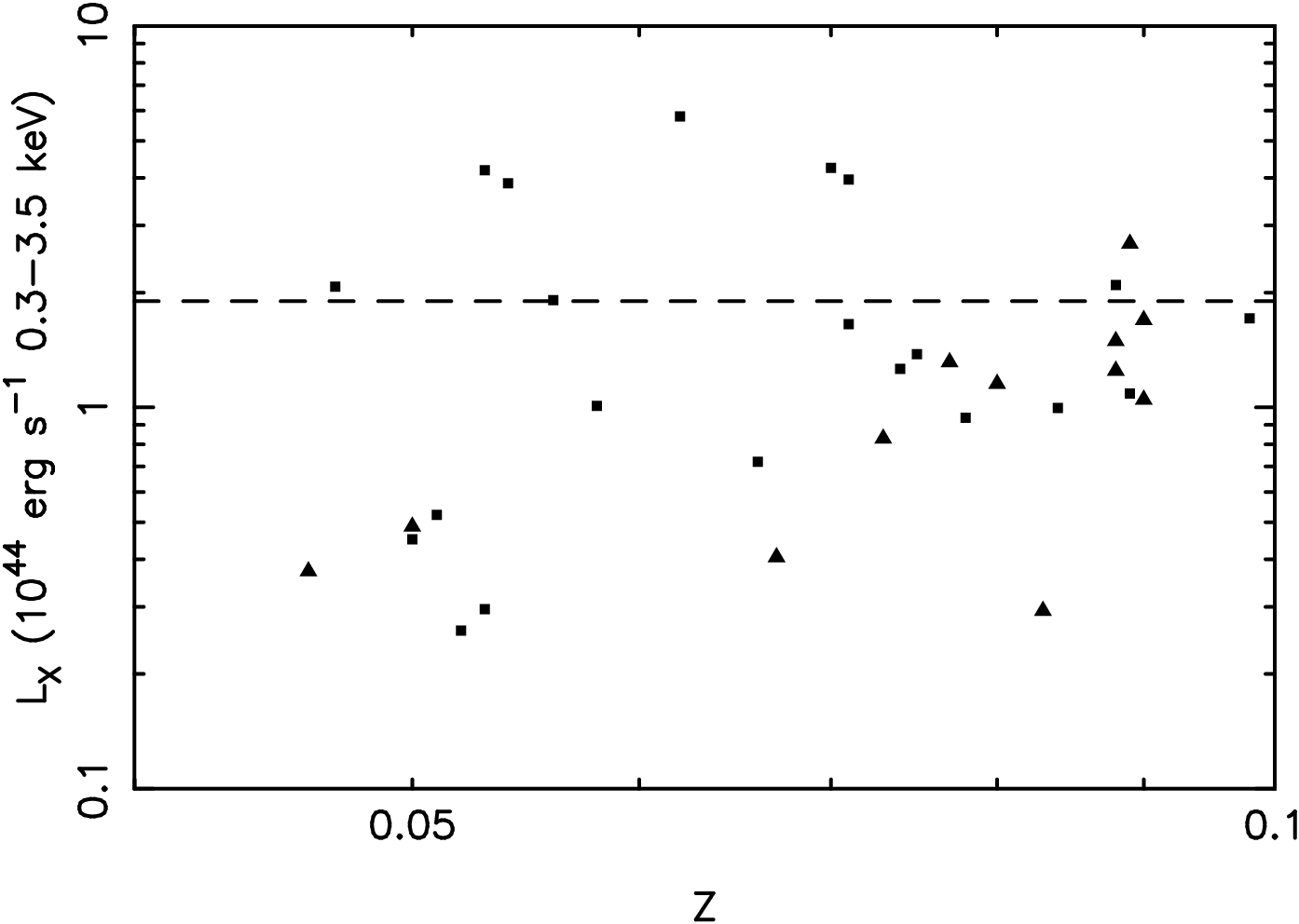}
}
}
\end{center}
\caption{The distribution of cluster X-ray luminosity with redshift for 
the host clusters of the 33 BCGs studied here.  The squares indicate 
clusters selected from the NORAS X-ray cluster catalogue \citep{boh00}, 
the triangles indicate those selected from the EMSS X-ray cluster catalogue 
\citep{gl94}.  The dashed line indicates a cluster X-ray luminosity of 
$L_X = 1.9 \times10^{44}$erg s$^{-1}$, defined as marking the division
between the two populations of BCGs at redshifts $z>0.1$.}
\label{lx_distrib}
\end{figure}

The effects of masking companion galaxies and excluding points with
large errors in their intensities mean that
it was not possible to calculate $r_{\eta=1.5}$ for 2 out of the 35
BCGs in this sample, both in clusters with X-ray luminosities
$<1.9\times10^{44}$ erg s$^{-1}$.  This leaves a total sample of 33
BCGs with $0.04<z<0.1$ in clusters with X-ray luminosities
$0.26\times10^{44}<L_X$ erg s$^{-1}< 5.8\times10^{44}$.  The
distribution of the X-ray luminosities of the host clusters of these
33 BCGs is shown in Figure~\ref{lx_distrib}.  This illustrates both
the range of host cluster environment covered, over one order of
magnitude in X-ray luminosity, and that this range remains consistent
over the redshift range studied.  The Petrosian structural parameters
for these 33 BCGs are given in Table~\ref{struc_table_eta}.

\begin{table*}
\begin{center}
\caption{The Petrosian structural properties. 
$\langle \mu_{\eta} \rangle$ is corrected for K-dimming, cosmological
dimming and Galactic absorption as described in the text.}
\begin{tabular}{||c|c|c|c|c|c|c|c||}
\hline
Cluster & z & $L_X$ & $r_{\eta=1.5}$ & $\Delta r_{\eta=1.5}$ & $r_{\eta=1.5}$ & $\langle \mu_{\eta=1.5} \rangle$ & $\Delta \langle \mu_{\eta=1.5} \rangle$\\
&&$\times10^{44}$erg s$^{-1}$&$h^{-1}$ kpc&$h^{-1}$ kpc&arcsec&mag arcsec$^{-2}$&mag arcsec$^{-2}$
\\ \hline
A133          &  0.056 &  1.912 &  15.53 &  1.85 & 20.20 &  18.64 &  0.21  \\ 
A292          &  0.066 &  0.720 &  16.26 &  1.87 & 18.22 &  18.58 &  0.21  \\ 
A399          &  0.071 &  3.960 &  26.17 &  3.17 & 27.43 &  19.38 &  0.27  \\ 
A401          &  0.074 &  1.263 &  27.07 &  3.15 & 27.22 &  19.72 &  0.21  \\ 
A478          &  0.088 &  2.095 &  19.57 &  2.19 & 16.97 &  19.57 &  0.21  \\ 
A566          &  0.098 &  1.714 &  13.62 &  1.88 & 10.71 &  19.13 &  0.25  \\ 
A644          &  0.070 &  4.246 &  21.84 &  2.70 & 23.22 &  19.90 &  0.24  \\ 
A671          &  0.050 &  0.451 &  19.65 &  2.76 & 28.60 &  19.30 &  0.24  \\ 
A754          &  0.053 &  4.190 &  16.66 &  1.87 & 22.88 &  18.53 &  0.20  \\ 
A757          &  0.051 &  0.523 &   5.96 &  0.59 &  8.46 &  18.34 &  0.17  \\ 
A763          &  0.084 &  0.996 &   8.84 &  1.32 &  7.96 &  19.15 &  0.31  \\ 
A780          &  0.054 &  3.870 &  14.72 &  1.41 & 20.02 &  19.28 &  0.16  \\ 
A970          &  0.058 &  1.009 &   5.98 &  1.01 &  7.52 &  17.99 &  0.34  \\ 
A978          &  0.052 &  0.260 &  12.62 &  2.39 & 17.55 &  18.72 &  0.37  \\ 
A1291         &  0.053 &  0.296 &  16.85 &  1.59 & 23.12 &  20.49 &  0.26  \\ 
A1541         &  0.089 &  1.088 &  24.38 &  3.18 & 20.91 &  19.84 &  0.27  \\ 
A1644         &  0.047 &  2.073 &  23.58 &  2.21 & 36.32 &  19.63 &  0.16  \\ 
A1775         &  0.071 &  1.653 &  29.27 &  2.84 & 30.61 &  20.20 &  0.26  \\ 
A1795         &  0.062 &  5.797 &  22.80 &  3.42 & 27.17 &  20.06 &  0.34  \\ 
A1800         &  0.075 &  1.379 &  18.94 &  2.06 & 18.89 &  18.76 &  0.19  \\ 
A1809         &  0.078 &  0.940 &  12.86 &  1.71 & 12.32 &  18.65 &  0.24  \\ 
MS0007-2      &  0.050 &  0.487 &  11.62 &  1.50 & 16.98 &  18.58 &  0.24  \\ 
MS0102-3      &  0.080 &  1.154 &  10.35 &  1.55 &  9.79 &  18.11 &  0.29  \\ 
MS0301-7      &  0.083 &  0.293 &  10.50 &  1.78 &  9.60 &  18.91 &  0.29  \\ 
MS0904-5      &  0.073 &  0.829 &  11.76 &  1.63 & 12.09 &  18.32 &  0.26  \\ 
MS1306-7      &  0.088 &  1.494 &  15.63 &  4.01 & 13.68 &  18.64 &  0.30  \\ 
MS1531-2      &  0.067 &  0.405 &   8.06 &  1.00 &  8.98 &  18.44 &  0.21  \\ 
MS1558-5      &  0.088 &  1.249 &  19.19 &  1.62 & 16.66 &  19.38 &  0.16  \\ 
MS1754-9      &  0.077 &  1.315 &  10.36 &  1.75 & 10.14 &  17.97 &  0.36  \\ 
MS2215-7      &  0.090 &  1.049 &  14.42 &  2.27 & 12.27 &  19.32 &  0.32  \\ 
MS2216-0      &  0.090 &  1.697 &  21.65 &  2.34 & 18.42 &  19.39 &  0.22  \\ 
MS2318-7      &  0.089 &  2.688 &  19.40 &  1.66 & 16.67 &  18.95 &  0.16  \\ 
MS2354-4      &  0.046 &  0.372 &  13.06 &  0.85 & 20.65 &  18.41 &  0.11 \\ \hline
\label{struc_table_eta}
\end{tabular} 
\end{center}
\end{table*}

Statistical errors on the measurement of $r_{\eta}$ and $\langle
\mu_{\eta} \rangle$ were calculated by Monte Carlo realisations of the
$\eta$ profile.  The profile was resampled from a series of Gaussian
profiles each with a FWHM equal to the rms of each point, and the
parameters measured for each resampled profile.  This process was
repeated $1000$ times resulting in 1000 independent measurements of
each parameter, and, therefore, the spread around them.  The standard
deviation around the resampled parameters is of the order of 10 per
cent in $r_{\eta}$ and $\Delta \langle \mu_{\eta}
\rangle \sim0.4$ mag arcsec$^{-2}$.

\subsection{Consistency}
\label{struc:consistent}
It is possible to check the consistency of the analysis methods described 
above as repeat observations were made with both telescopes.  

Table~\ref{etarepeat} summarises the Petrosian parameters measured for
the three BCGs with repeat observations made by UH and the two BCGs
with repeat observations made by UKIRT.  Two galaxies (those in the
clusters MS0301-7 and MS0904-5) were observed by both UH and UKIRT,
enabling a comparison of the photometry between the two telescopes.
The 5 galaxies have a mean absolute difference $\bar{\Delta
r_{\eta}}=0.86\pm0.27 h^{-1}$ kpc and $\bar{\Delta \langle\mu_{\eta}
\rangle} =0.20\pm0.14$ mag arcsec$^{-2}$.  The errors quoted are the 
error on the mean $(=\sigma/\sqrt{N})$. The difference in  $\langle\mu_{\eta}
\rangle$ is comparable to the statistical errors, however the statistical 
error on $r_{\eta}$ is significantly larger.

\small
\begin{table*}
\begin{center}
\caption{Summary of the Petrosian parameters measured from repeat observations.  The upper table shows the Petrosian radii, $r_{\eta}$ ($h^{-1}$ kpc), calculated for each night, the lower table shows the mean Petrosian surface brightness, $\langle \mu_{\eta} \rangle$ (mag arcsec$^{-2}$).}
\label{etarepeat}
\begin{tabular}{|c|c|c|c|c|c|c|c|}
\hline
Cluster&\multicolumn{6}{c|}{$r_{\eta}$ ($h^{-1}$ kpc)}&Mean\\ 
&8 Nov 1994&9 Nov 1994&11 Jan 1997&14 Jan 2001&15 Jan 2001&16 Jan
2001&\\ \hline 
A399&&&&25.99&26.34&&26.17\\
A978&&&&12.48&&12.76&12.62\\ 
MS0301-7&11.51&&&&&9.48&10.50\\
MS0904-5&&11.57&12.47&&11.09&11.89&11.76\\
MS1306-7&&15.74&15.52&&&&15.63\\ \hline 
&&&&&\multicolumn{2}{c}{Mean
Absolute Difference:}&0.86$\pm0.27$\\
\hline
\multicolumn{8}{c}{ }\\
\hline
Cluster&\multicolumn{6}{c|}{$\langle \mu_{\eta} \rangle$ (mag arcsec$^{-2}$)}&Mean\\
&8 Nov 1994&9 Nov 1994&11 Jan 1997&14 Jan 2001&15 Jan 2001&16 Jan 2001&\\ \hline
A399&&&&19.43&19.43&&19.43\\
A978&&&&19.02&&18.48&18.75\\
MS0301-7&18.86&&&&&19.09&18.98\\
MS0904-5&&18.35&18.46&&18.29&18.38&18.40\\ 
MS1306-7&&18.72&18.69&&&&18.70\\ \hline
&&&&&\multicolumn{2}{c}{Mean Absolute Difference:}&0.20$\pm$0.14\\ \hline
\end{tabular} 
\end{center}
\end{table*}
\normalsize

These comparisons confirm that there are no significant offsets between the 
observations made on different nights or with the different telescopes.  This 
also demonstrates the stability of the parameters fitted and the validity of 
the analysis techniques used.

\section{Results} 
\label{pet_kormendy}

Two of the BCGs in Figure~\ref{eta_profile} (those in Abell 644 and
Abell 1644) clearly show a plateau in their $\eta$ profiles.  This has
been observed previously by \cite{kjaergaard93} in NGC 4874, one of
the two BCGs in Coma, which is known to have a cD morphology.  This
plateau is not present in the profiles of normal cluster elliptical
galaxies that follow perfect de Vaucouleurs laws
(Figure~\ref{kjaer_eta};
\citealt{sandage90,sandage90b,sandage91,kjaergaard93}).  Abell 644 and 
Abell 1644 have been classified as cD clusters by \cite{struble87},
however, the BCGs themselves have not yet been individually classified
as cD galaxies.  The plateaux we observe are analagous to those
observed by \cite{kjaergaard93} in NGC 4874.  We therefore speculate
that the plateaux are likely to be a signature of the cD morphology.
The detailed study of the profiles of cD galaxies by \cite{schombert3}
suggests that the envelopes of cD galaxies become dominant over the
underlying galaxy component at an absolute surface brightness $\sim
\mu_V\geq24$ mag arcsec$^{-2}$, this is evident as a break in the 
surface brightness profile at these surface brightnesses.
Assuming a $V-K$ colour $\sim3.0$ mag from \cite{ble92}, this
corresponds to $\mu_K\geq21$ mag arcsec$^{-2}$ for these data.  The
plateaux in these data begin at $r\sim10~h^{-1}$ kpc and $\mu_K
\sim19$ mag arcsec$^{-2}$ suggesting that the halo of BCGs with cD
morphologies is distinguishable at higher surface brightnesses than
found previously.

The form of the Petrosian profile indicates a possible fundamental
difference between galaxies with cD morphologies and those that follow
de Vaucouleurs profiles.  The signal-to-noise ratio of these data
preclude further analysis of BCGs with cD morphologies, and it is not
studied further.  However, it is noted that the Petrosian profile
could provide an objective means with which to determine cD morphology
without obtaining very deep imaging or fitting empirical models, which
do not provide an adequate fit (Figure~\ref{eta_profile},
Figure~\ref{kjaer_eta}; \citealt{graham96}).  However, this requires
further examination.



\begin{figure}
\begin{center}

    \resizebox{20pc}{!}{
      \rotatebox{-90}{
        \includegraphics{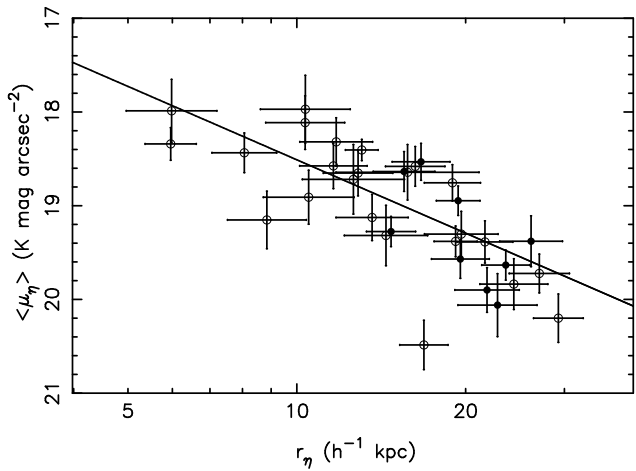}
      }
    }

  \end{center}
\caption{The Petrosian Kormendy relation.  The error bars indicate 
the $1\sigma$ statistical errors on the measured parameters.  The open
points denote BCGs in clusters with $L_X (0.3-3.5$
keV$)<1.9\times10^{44}$erg s$^{-1}$ whilst the filled points denote
BCGs in clusters with $L_X (0.3-3.5$ keV$)>1.9\times10^{44}$erg
s$^{-1}$. The solid line marks the best-fit relation described by
Equation~\ref{eq:pet_kor_lo}.}
\label{loz_petkor}
\end{figure}

The Petrosian Kormendy relationship is shown for all 33 galaxies in
Figure~\ref{loz_petkor}.  The error bars show the statistical errors
on these measurements.  To illustrate any environmental dependence of
BCG properties on the Kormendy relation, the cluster X-ray luminosity,
$L_X = 1.9 \times10^{44}$erg s$^{-1}$ in the EMSS (0.3-3.5 keV)
passband, defined in \cite{brough02} as marking the division between
the two populations of BCGs at redshifts $z>0.1$, is used.  There is
an offset between the properties of the BCGs on this relation,
depending on their host cluster environment, but they both follow the
same relationship.
The best-fit Kormendy relation is given by:
\begin{equation}
\langle \mu_{\eta} \rangle=2.60^{(0.03)}~{\rm log}_{10}~r_{\eta}-16.07^{(0.6)}.
\label{eq:pet_kor_lo}
\end{equation}
The quantities in brackets are the standard deviations around the
gradient and intercept solutions.  The scatter around the relation,
$\sigma=0.42$ mag arcsec$^{-2}$.

The Kormendy relations fitted to other samples of Petrosian structural
parameters are summarised in Table~\ref{kor_table}.  This sample
follows a similar gradient to those fitted by \cite{sandage91} and
\cite{kjaergaard93} to samples of BCGs, although it has a slightly
larger scatter with $\sigma=0.42$ mag arcsec$^{-2}$ in contrast to
$\sigma\sim0.3$ mag arcsec$^{-2}$.

\begin{table}
\begin{center}
\caption{Kormendy relations fitted to BCG Petrosian structural parameters, the relationships are of the form $\langle\mu_{\eta}\rangle=M\times {\rm log_{10}}r_{\eta} + C$.}
\label{kor_table}
\begin{tabular}{|c|c|c|c|c|c|}
\hline
Author&N&$\lambda$&M&C&$\sigma$\\ \hline
\cite{sandage91}&56&V&3.12&8.51&0.27\\ 
\cite{kjaergaard93}&18&V&2.11&18.23&0.31\\ 
\cite{kjaergaard93}&18&B&2.09&19.45&0.29\\ 
This sample &33&K&2.60&15.9&0.42\\ \hline
\end{tabular} 
\end{center}
\end{table}

To examine the environmental dependence of these parameters further
the means and dispersions (split by the X-ray luminosity of their
host cluster) are summarised in Table~\ref{petr_table}.

\begin{table}
\begin{center}
\caption{Statistical properties of the BCG Petrosian effective radii, 
$r_{\eta}$ ($h^{-1}$ kpc), and mean surface brightness, $\langle
\mu_{\eta} \rangle$ ($K$ mag arcsec$^{-2}$), as a function of cluster
X-ray luminosity, $L_X$.}
\label{petr_table}
\begin{tabular}{|c|ccc|}
\hline
$L_{{\rm X}}(\times10^{44}$erg s$^{-1})$&N&$r_{\eta}~(1 \sigma)$&$\langle \mu_{\eta} \rangle~(1 \sigma)$\\ \hline
$<1.9$&24&14.96 (6.23) &18.93 (0.68)\\ 
$>1.9$& 9&20.03 (3.90) &19.33 (0.53)\\ 
\hline
\end{tabular} 
\end{center}
\end{table}

\begin{figure}
\begin{center}

    \resizebox{20pc}{!}{
      \rotatebox{-90}{
        \includegraphics{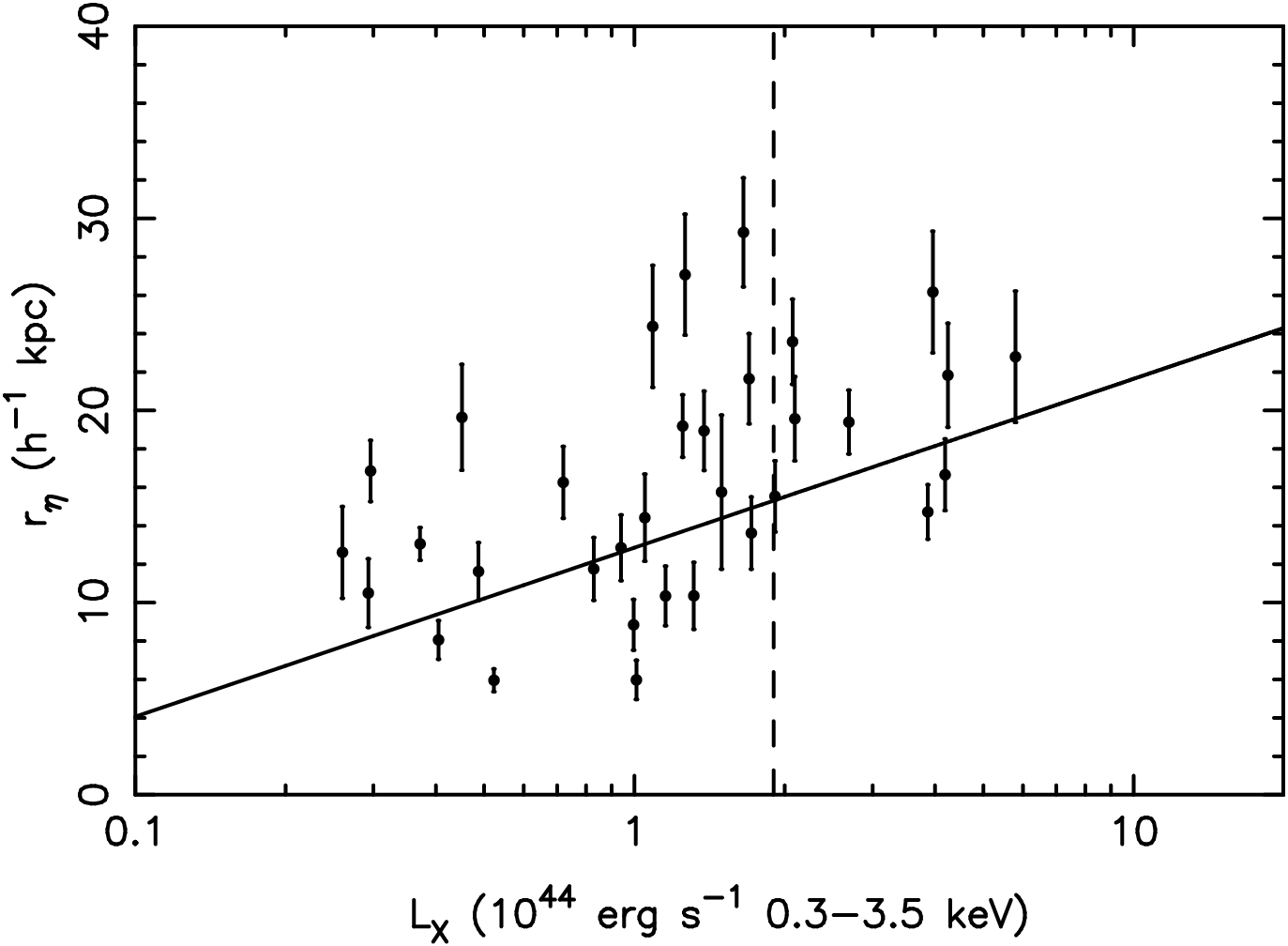}
	\includegraphics{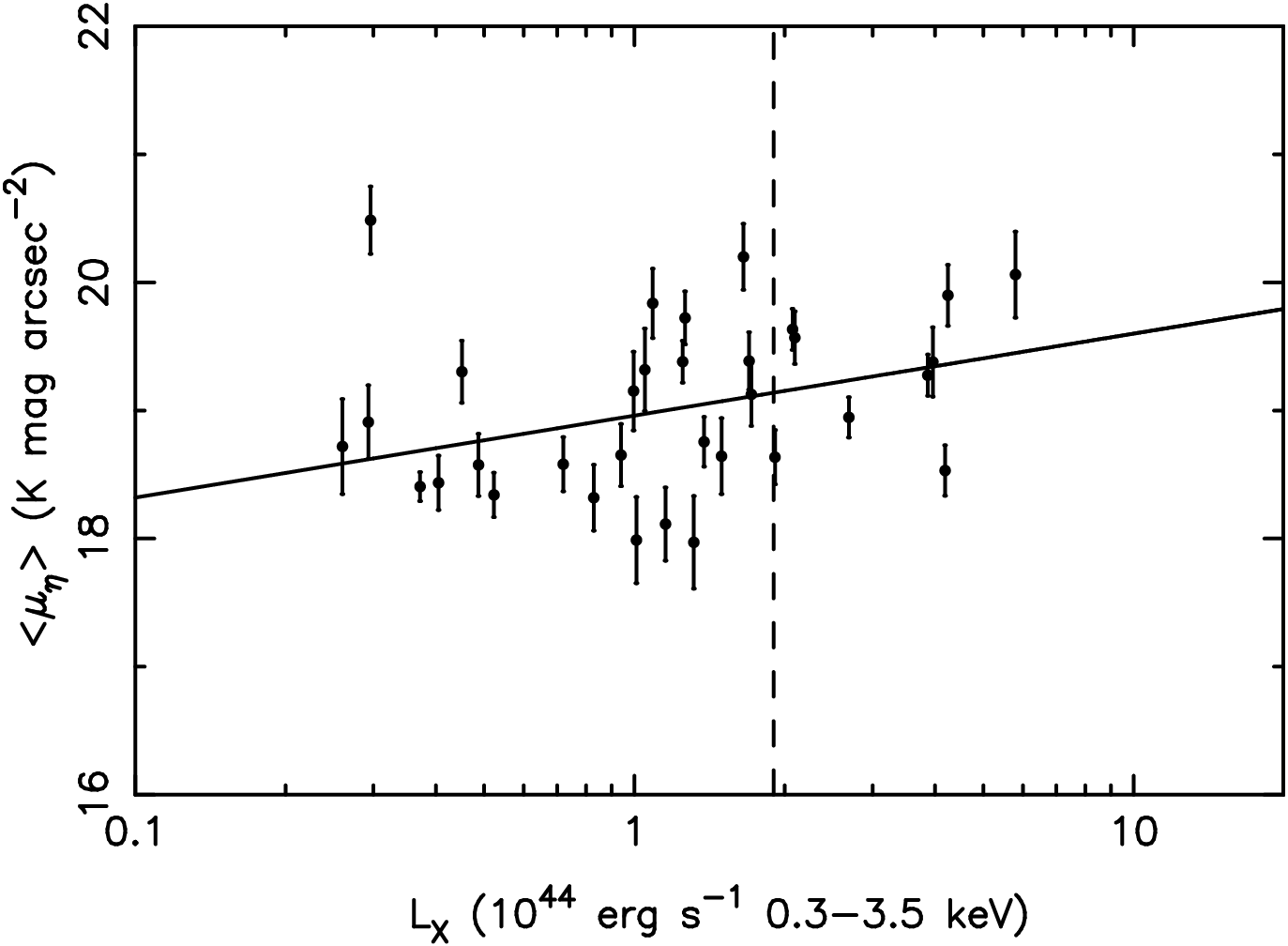}
      }
    }

  \end{center}
\caption{The distribution of Petrosian radius, $r_{\eta}$ (upper panel), 
and mean surface brightness $\langle \mu_{\eta} \rangle$ (lower
panel), with cluster X-ray luminosity, $L_X$, for 33 BCGs. The error
bars indicate the $1\sigma$ statistical errors on the parameters.  The
dashed line indicates the empirical $L_X=1.9\times10^{44}$ erg
s$^{-1}$ division between high and low X-ray luminosity clusters.  The
solid line in the upper panel indicates the straight-line fit given by
Equation~\ref{eq:env_corr}, Equation~\ref{eq:mu_corr} is illustrated
by the solid line in the lower panel.  }
\label{lx_eta}
\end{figure}

Figure~\ref{lx_eta} illustrates the distributions of the BCG Petrosian
radii (upper panel) and surface brightnesses (lower panel) as a
function of their host cluster environment.  
A Spearman rank correlation on the Petrosian radii confirms that the
BCG radii are correlated with their host cluster environment, with a
correlation coefficient of $r_s=0.55$.  For 33 galaxies, the Student's
t-test rejects the null hypothesis that there is no correlation
at $>99$ per cent confidence level.  A weighted straight-line fit
takes the statistical errors in $r_{\eta}$ into consideration and also
results in a statistically significant correlation between $r_{\eta}$
and $L_X$:
\begin{equation}
r_{\eta}=8.78^{(0.16)}{\rm log_{10}}~L_X+12.9^{(2.9)}.
\label{eq:env_corr}
\end{equation}


This is not a selection effect, Figure~\ref{lx_distrib} illustrates
that the host clusters of the BCGs have a uniform distribution in
X-ray luminosity over the redshift range of the sample and
Section~\ref{struc:consistent} demonstrates that there are no offsets
between the parameters measured at the different telescopes.  The BCG
radii are clearly correlated with their host cluster environment.

In contrast, the lower panel of Figure~\ref{lx_eta} indicates that the
Petrosian mean surface brightnesses are not strongly correlated with
their environment.
A Spearman rank correlation suggests that the Petrosian surface
brightnesses are correlated with their environment with a correlation
coefficient $r_s=0.36$.  The Student's t-test rejects the null
hypothesis of no correlation at $>95$ and $<98$ per cent confidence
level, suggesting that there is a correlation of $\langle \mu_{\eta}
\rangle$ with environment.  A weighted straight-line fit demonstrates
that the mean surface brightnesses of BCGs are weakly correlated with
their host cluster environment.
\begin{equation}
\langle \mu_{\eta} \rangle=0.64^{(0.02)}{\rm log_{10}}~L_X+19.0^{(0.3)}.
\label{eq:mu_corr}
\end{equation}

The observed relationships between BCG parameters and those of their
host cluster could be due to a systematic error in the background
subtraction of these galaxies.  In order to test this, the
$1\sigma_{background}$ calculated by SExtractor was used to model the
possible systematic errors.  This value is typically $\sim2$
counts/pixel and is significantly more than the differences between
the two background subtraction methods compared in
Section~\ref{analysis_4}.  The images were reanalysed -- first with an
extra $1\sigma_{background}$
subtracted and again with $1\sigma_{background}$ added.

The Petrosian profiles of galaxies in those images with
$1\sigma_{background}$ added fall off to zero before reaching
$\eta=1.5$ and it is not possible to measure their structural
parameters.  This is not seen in the original data and is illustrative
that the systematic errors are not this large.

\begin{figure}
\begin{center}

    \resizebox{20pc}{!}{
      \rotatebox{-90}{
        \includegraphics{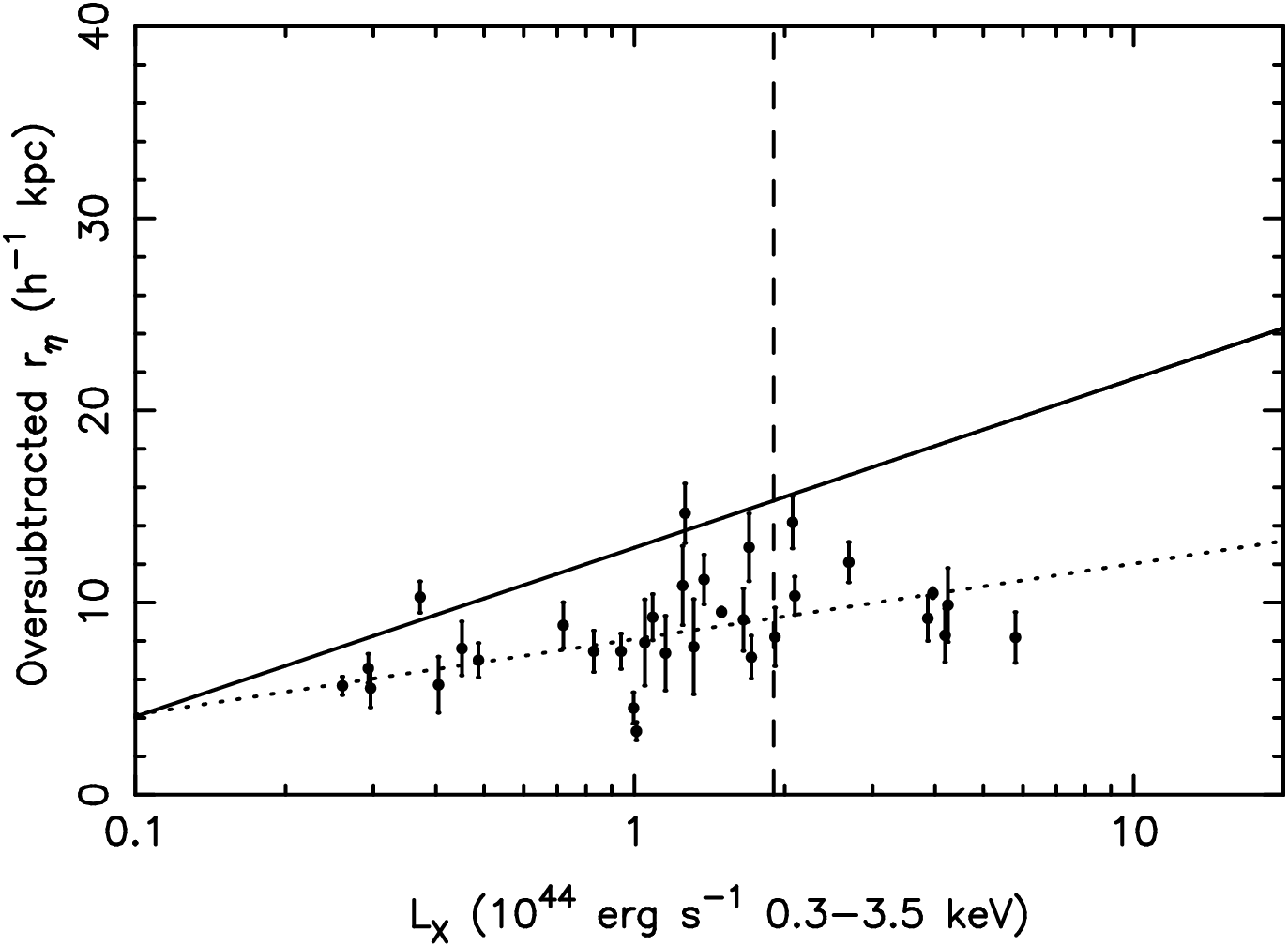}
	\includegraphics{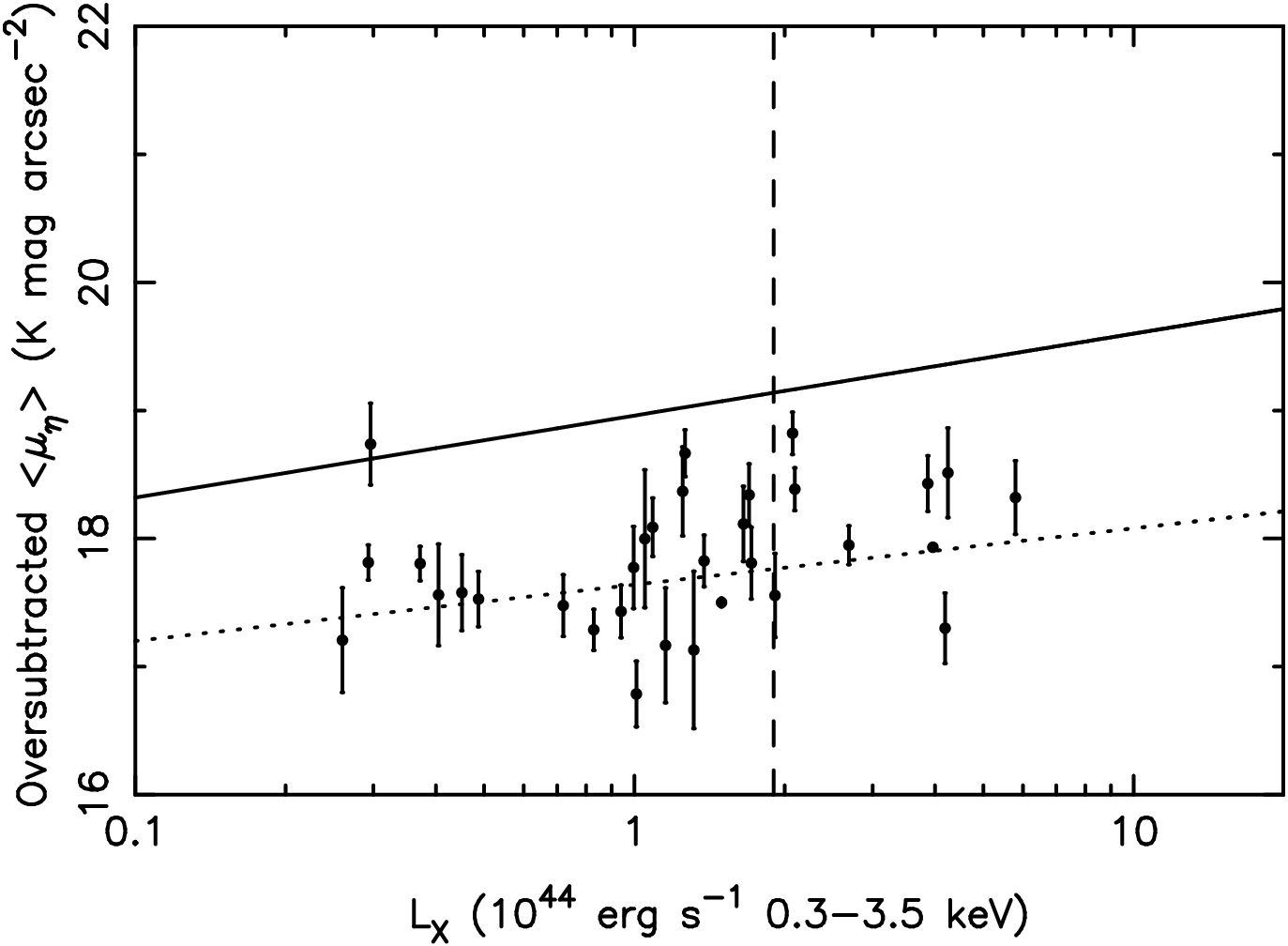}
      }
    }

  \end{center}
\caption{The distribution of Petrosian parameters $r_{\eta}$ (upper 
panel), and mean surface brightness $\langle \mu_{\eta} \rangle$
(lower panel) measured from images with the background oversubtracted
by $1\sigma_{background}$. The error bars indicate the $1\sigma$
statistical errors on the parameters.  The dashed line indicates the
empirical $L_X=1.9\times10^{44}$ erg s$^{-1}$ division between high
and low X-ray luminosity clusters.  The solid line in each panel gives
the straight-line fit fitted to the original data, and the dotted line
indicates the straight-line fit to the oversubtracted data, described
by Equation~\ref{eq:backre_corr} (upper panel) and
Equation~\ref{eq:backmue_corr} (lower panel).}
\label{backtest_lx_eta}
\end{figure}

The Petrosian radii of BCGs in images with $1\sigma_{background}$
subtracted are smaller and the mean surface brightnesses are brighter
than the original measurements.  This is illustrated in
Figure~\ref{backtest_lx_eta}, which also demonstrates that, despite
the significant background oversubtraction, the relationships with the
Petrosian structural parameters and the X-ray luminosities of their
host cluster environments remain.  The weighted straight-line fit to
the upper panel of Figure~\ref{backtest_lx_eta}:
\begin{equation}
r_{\eta}=3.93^{(0.06)}{\rm log_{10}}~L_X+8.1^{(0.6)}.
\label{eq:backre_corr}
\end{equation}
The weighted straight-line fit to the lower panel of Figure~\ref{backtest_lx_eta}:
\begin{equation}
\langle \mu_{\eta} \rangle=0.44^{(0.01)}{\rm log_{10}}~L_X+17.6^{(0.2)}.
\label{eq:backmue_corr}
\end{equation}

The BCG parameters are clearly correlated with those of their host
cluster environment.

The possibility that the observed relationships between the BCG
parameters and those of their environment are due to a systematic
error in the background subtraction of these galaxies was further
tested by a comparison with \cite{garilli97}
These authors examined BCGs in the $r$-band 
which is less susceptible to problems with background subtraction than
the near-infrared, we have 7 BCGs in common.

\cite{garilli97} 
fit de Vaucouleurs profiles to their BCGs and quote the fitted
effective radius, $r_e$, and effective surface brightness, $\mu_e$.
Using these values, it is possible to construct the equivalent
Petrosian profile for that de Vaucouleurs profile (as seen in
Figure~\ref{kjaer_eta}).  First, the effective radius, $r_e$ was
transformed into the cosmology used here.
The Petrosian radius was then measured at $\eta=1.5$ as described
above.  \cite{garilli97} 
do not circularize their profiles as we have done here so these
comparisons are made with the major-axis profiles of these galaxies.
It was not, therefore, possible to compare $r_\eta$(this work) with
$r_\eta$(comparison) for the BCG in MS1558-5 as the major-axis profile
for this galaxy does not reach $\eta=1.5$.  This leaves a sample of 6
galaxies in common.  The measured values are given in
Table~\ref{comp_table}.  We also note that the $r_{\eta}$ of the BCG
in Abell 1644 measured by S. Andreon is in apparent agreement with
this work (S. Andreon, private communication).

The mean difference $r_\eta$(this work)$-r_\eta$(comparison)
$=-9.3\pm5.2~h^{-1}$ kpc.  The most discrepant galaxy is the BCG in
Abell 401, with $r_\eta$(this work)$-r_\eta$(comparison)
$=33.92~h^{-1}$ kpc.  However, \cite{garilli97} note that they
determine a de Vaucouleurs effective radius for this BCG which is
$\sim25$ per cent larger than that obtained by \cite{malamuth85}
($r_e=91.04$ kpc).  If we discount this galaxy the mean difference
between the two samples is then $-4.3\pm1.9~h^{-1}$ kpc.  This
difference is of the order of the errors on $r_\eta$(this work) which
are $\Delta r_\eta\sim4~h^{-1}$ kpc (Table~\ref{comp_table}).
We therefore conclude that the observed relationships are not due to a
systematic error in the background calculation of these galaxies.


%



\begin{table}
\begin{center}
\caption{Comparing major-axis $r_\eta$ measured from de Vaucouleurs 
profiles fitted by Garilli et al. (1997) with galaxies in common with
this work.}
\label{comp_table}
\begin{tabular}{ccccc}
\hline
Cluster&$r_e$&$r_\eta$ (comp)&$r_\eta$ (this work)&$\Delta r_\eta$\\
&$h^{-1}$ kpc&$h^{-1}$ kpc&$h^{-1}$ kpc&$h^{-1}$ kpc\\
\hline
A399      &  35.12  & 44.67 & 36.20 & 11.79\\
A401      &  59.67  & 74.67 & 40.75 & 6.47\\
A671      &  21.74   & 28.28 & 23.92 & 5.67\\
MS0102-3  &  9.41   & 13.01 & 13.97 & 3.26\\
MS0301-7  &  10.99   & 15.05 & 13.92 & 3.26\\
MS0904-5  &  15.66   & 20.80 & 12.24 & 2.61\\
\hline
\end{tabular} 
\end{center}
\end{table}


\section{Discussion}
\label{discuss_4}

In order to understand what the observed correlation between Petrosian
radius and host cluster environment means in terms of BCG formation
and evolution, the effect that mergers have on the surface brightness
profiles of galaxies is considered.

As introduced in Section~\ref{intro}, equal-mass, dissipationless mergers of 
simulated elliptical galaxies produce structurally non-homologous remnants 
and are, therefore, able to reproduce the observed FP of elliptical galaxies 
\citep{capelato95,nipoti03}.  However, these simulations are unable to 
reproduce the projections of the FP, the Faber-Jackson and Kormendy
relations.  The Faber-Jackson relation is characterized by a lower
velocity dispersion than predicted by the Faber-Jackson relation for a
given luminosity increase and the Kormendy relation by significantly
larger radii than predicted from observations
\citep{navarro90,ciottivA01,nipoti03}.  These simulations are,
however, consistent with observations of BCGs.  BCGs also lie on the
FP for normal cluster elliptical galaxies \citep{bcgfp} but are
observed to have significantly lower velocity dispersions than
predicted by the Faber-Jackson relation for normal cluster elliptical
galaxies \citep{malamuth81,malamuth85,bcgfp} and larger radii than
predicted by the normal cluster elliptical galaxy Kormendy relation
\citep{thuan81,hoessel87,schombert2,bcgfp}.  This suggests that BCGs
can form from equal-mass, dissipationless mergers of large luminous
elliptical galaxies.  However, gas dissipation must be important in
the formation of less massive elliptical galaxies in order to increase
their central velocities, consistent with the observed Faber-Jackson
and Kormendy relations.

Numerical simulations of elliptical galaxy mergers indicate that the
effective radii of merger remnants increase proportionally with total
mass, by a factor of $1-2$
\citep{h80,navarro90,capaccioli92,capelato95,nipoti03}.  There is a
range of increase factors because the properties of the remnant are
dependent on the energy and angular momentum of the collision as well
as the mass ratios of the progenitors.  Head-on, equal-mass
dissipationless mergers produce the largest remnants, with effective
radii double those of their progenitors
(e.g. \citealt{capelato95,nipoti03}).  From the predictions of current
simulations, the observation that BCGs in high X-ray luminosity
clusters are larger, by a factor of $\sim2$, than those in low X-ray
luminosity clusters at redshifts $z\leq0.1$ then suggests that these
galaxies have undergone one or two major (equivalent mass) merger
events, or several accretion events, more than those BCGs in low-$L_X$
clusters.  Figure~\ref{eta_profile} indicates that BCGs with cD
morphologies have larger Petrosian radii than would be measured if
they followed perfect de Vaucouleurs profiles.  Therefore, the BCG
radius -- cluster X-ray luminosity relationship could also suggest
that there are a higher proportion of BCGs with cD morphologies in
high X-ray luminosity clusters than there are in low X-ray luminosity
clusters.  Higher resolution images would be required to determine
this in a non-subjective manner.

Numerical simulations indicate that the density profiles of the
remnant galaxies remain the same as those of their progenitors,
therefore, the effective surface density should scale as $I_e\propto
M_r/r_r^2$ \citep{navarro90}.  In magnitudes, this corresponds to a
change in $\langle \mu_e \rangle$ of up to 1 mag arcsec$^{-2}$ in 2
consecutive equal-mass mergers between elliptical galaxies, as
predicted by the simulations of \cite{capelato95} and \cite{evstigneeva04}.  
The corresponding change in surface brightness predicted is within the
observational scatter of this data, explaining the lack of a clear
correlation of BCG mean surface brightness with host cluster
environment.

\cite{brough02} concluded that BCGs in high X-ray luminosity clusters 
showed little sign of an increase in mass since $z\sim1$, in contrast
to those in low X-ray luminosity clusters which have increased in mass
by up to a factor of four over the same timescale.  Combining the
results presented here with those of \cite{brough02} a consistent
picture emerges.  These two observations suggest that BCGs in high
X-ray luminosity clusters assembled their stellar mass at redshifts
$>1$ and have been passively evolving since, in contrast to those BCGs
in low X-ray luminosity clusters which are still in the process of
assembling today.  This is qualitatively consistent with the theories
of hierarchical structure formation which suggest that the less
massive systems (clusters and galaxies) we observe today assembled
more recently than the most massive systems (e.g. \citealt{kc98}).

\section{Conclusions}
\label{summary_4}

In this paper, the Petrosian parameters measured from the surface
brightness profiles of 33 BCGs in clusters covering a wide range in
environmental density have been presented, with the aim of
interpresting their evolutionary history.

Analysis of the Petrosian $\eta$ profiles of this sample indicates
that there is a deviation of the Petrosian profiles of some BCGs with
cD morphologies from those of a galaxy following a perfect de
Vaucouleurs law.  This is suggested to be a signature of galaxies with
cD morphologies and may indicate a fundamental structural difference.

  
The structural properties of BCGs are clearly determined by their host
cluster environment, as suggested by \cite{garilli97}.  BCGs in high
X-ray luminosity clusters are larger and fainter than those in
low X-ray luminosity clusters.  

Research into other cluster galaxy populations suggests that their
properties are driven, not by the global cluster environment, but by
their local environment as measured by the density of their nearest
neighbours, with those galaxies in the least dense environments,
showing the most signs of evolution.  The unique position of BCGs
means that their local environment {\it is} the global cluster
environment and this affects their evolution such that those BCGs in
the least dense environments have undergone more evolution since
$z\sim1$ than those in the most dense clusters.  This provides an
indication of how a single population of galaxies evolves
hierarchically.  It is likely that BCGs arise from the merging of
satellites, initially drawn towards the cluster centre by collimated
infall along filaments in the early epochs of cluster formation,
however, this work suggests that this process is continuing in the
least dense clusters at the present time.

\section*{Acknowledgments}
We thank the referee Stefano Andreon for his valuable
suggestions.  SB acknowledges PPARC for a Postgraduate Studentship
and Phil James, Alfonso Arag\'{o}n-Salamanca and Dave Carter for
helpful comments.  DJB acknowledges the support of NASA contract
NAS8-39073.  We also acknowledge the use of the UH 2.2m telescope at
the Mauna Kea Observatory, Institute for Astronomy, University of
Hawaii.

\bsp

\label{lastpage}

\end{document}